\documentclass[11pt]{article}
\usepackage{amssymb}
\usepackage{graphics}
\usepackage{epsfig}
\usepackage{a4wide}

\begin{document}
\newcommand{\rrttbb}{$\gamma\gamma \to t\bar t b\bar b$ }

\title{QCD corrections to $t\bar t b \bar b$ productions via photon-photon
collisions at linear colliders}
\author{ Guo Lei, Ma Wen-Gan, Han Liang, Zhang Ren-You, and Jiang Yi  \\
{\small Department of Modern Physics, University of Science and Technology}\\
{\small of China (USTC), Hefei, Anhui 230027, P.R.China}  }

\date{}
\maketitle \vskip 15mm
\begin{abstract}
We calculated the complete next-to-leading order(NLO) QCD
corrections to the $t\bar t b \bar b$ production process at a
$\gamma \gamma$ collider in the standard model($SM$). The
calculation of the one-loop QCD correction includes the
evaluations of the hexagon and pentagon amplitudes. We studied the
NLO QCD corrected total cross section, the distributions of
transverse momenta of final top- and bottom-quark states, and the
dependence of the cross section on renormalization scale $\mu$. It
shows that NLO QCD correction generally increases the LO cross
section in our chosen parameter space, and the K-factor varies
from $1.70$ to $1.14$ when colliding energy goes up from $400~GeV$
to $2~TeV$. We find that the correction distinctly changes the
distributions of transverse momenta of the final top- and
bottom-quark states, and the NLO QCD correction obviously improves
the independence of the cross section for process \rrttbb on the
renormalization scale.
\end{abstract}


\vfill \eject

\baselineskip=0.32in

\renewcommand{\theequation}{\arabic{section}.\arabic{equation}}
\renewcommand{\thesection}{\Roman{section}.}
\newcommand{\nb}{\nonumber}

\makeatletter      % '@' is now a normal "letter" for TeX
\@addtoreset{equation}{section}
\makeatother       % '@' is restored as a "non-letter" character for TeX

\section{Introduction}
\par
Recently, the report of the top-quark mass measurements from the
CDF and D0 experiments at Fermilab, presents the preliminary world
average mass of the top quark is $m_t=172.5 \pm 1.3(stat) \pm
1.9(syst)~GeV$, which corresponds to a $20\%$ precision
improvement relative to the previous combination\cite{tew}. Due to
the huge top-quark mass being much heavier than the $W$-boson,
before the top-quark hadronization it undergoes dominantly the
weak decay via $t\to W^+ b$\cite{bigi}, that has two important
consequences different with other quarks, the narrow resonance
around the energy $2m_t$ is absent and the perturbative QCD is
reliable to study all the threshold region. We believe the large
top mass value will open up new vistas of electroweak physics and
possibilities for probing the effects beyond the $SM$. For
example, since the Higgs mechanism in the $SM$ and other extended
models predicts that the strength of quark-Higgs Yukawa coupling
is proportional to the quark mass\cite{farrell}, one can measure
top-quark Yukawa couplings in high precision to probe the $SM$ or
discover new physics. There have been many works concerning the
study of the top-quark physics at colliders
\cite{fadin,strassler,kwong,sumino,yndurain}. All those indicate
that the precise study of the top physics is accessible. In the
proposed and planned experiments, such as at the CERN LHC and the
ILC, more interests are focused on the developing and
understanding top physics and top characteristics.

\par
The future $e^+ e^-$ linear collider(LC) not only can be designed
to operate in $e^+e^-$ collision mode, but also can be operated as
a $\gamma\gamma$ collider. This is achieved by using Compton
backscattered photons in the scattering of intense laser photons
on the initial $e^+e^-$ beams. The top-quark pair production at
photon-photon collider is also useful for top-quark physics study.
It has been found that the $\gamma\gamma \to t\bar t$ production
rate with high $\gamma\gamma$ colliding energy, is much larger
than that from the direct $e^+e^-\to t\bar t$ production due to
the s-channel suppression of later process
\cite{chatterjee,han,denner0}.

\par
At the future International $e^+e^-$ Linear Collider (ILC), the
colliding center-of-mass-system (c.m.s.) energy can be raised up
to $2~TeV$. At $TeV$ scale high colliding energy, the final states
will be very complex and expected to be in multi-particles or jets
with large production rates, and top and Higgs signatures
naturally are included in these processes. For these processes
with large cross sections, a leading order analysis is not
adequate to make detailed predictions for their cross sections. In
fact, the processes with multiple final particles are particularly
interesting, since such processes often proceed through one or
more resonances that subsequently decay, or they represent an
irreducible background to such resonance processes. For example,
the associated production $\gamma\gamma\to t \bar t H$ is an
important process in probing top-Higgs Yukawa coupling in high
precision, and was already studied in Ref.\cite{chen}. But after
Higgs boson and top-quark decay, the $t \bar t H$ associated
production process at $\gamma\gamma$ collider would have the same
final states ($W^+bW^-\bar b b\bar b$) as the process \rrttbb.
Therefore, it is important to emphasize that the ability to
distinguish top, Higgs boson or other new particle signatures at
linear colliders, crucially depends on the understanding of the
corresponding backgrounds of the corresponding processes with
multi-particle final states. In order to exert all the abilities
of future colliders, precise theoretical predictions including
higher order corrections to multi-particle production processes
are necessary.

\par
In previous work, people used "double-pole approximation" (DPA)
\cite{dpa} to handle the evaluations of complete one-loop strong
and electroweak calculation for process including four particle
final states. Recently, the methods for the calculation of scalar
and the tensor 6-point integral functions were
provided\cite{denner1,Binoth1}. With the approach provided in
Ref.\cite{denner1}, the complete electroweak corrections to the
$e^+e^-\to 4f$ processes, which are relevant for W-pair
production, was presented by A. Denner, S. Dittmaier, M. Roth,
L.H. Wieders, and the results were compared with those in
Ref.\cite{denner2} by using DPA method.

\par
In this paper we present the calculations of the cross sections of
the process $\gamma\gamma \to t\bar t b\bar b$ at the
leading-order(LO) (${\cal O}(\alpha_{ew}^2\alpha_s^2)$) and QCD
next-to-leading-order(NLO) (${\cal O}(\alpha_{ew}^2\alpha_s^3)$).
The paper is organized as follow: The tree-level analytical
calculation of the cross section for the process \rrttbb is given
in section II. In section III the analytical calculation of the
NLO QCD corrections is presented. The numerical results and
discussions are given in section IV. Finally, a short summary is
given in section V.

\vskip 10mm
\section{Analytical calculation of the cross section
for the process \rrttbb at the tree-level(${\cal
O}(\alpha_{ew}^2\alpha_s^2)$) }

\par
The process \rrttbb can be induced via $t-$ and $u-$channel at the
tree-level. In this section we consider the tree-level$({\cal
O}(\alpha_{ew}^2\alpha_s^2))$ contribution to the cross section
for the process \rrttbb. We denote the process as
\begin{equation}
\label{process} \gamma(p_1)+\gamma(p_2) \to t(p_3)+ \bar
t(p_4)+b(p_5)+\bar b(p_6),
\end{equation}
where the four-momenta of incoming photons are denoted as $p_1$
and $p_2$, and $p_{3}$, $p_{4}$, $p_{5}$, $p_{6}$ represent the
four-momenta of the final particles. The FeynArts3.2
package\cite{fey} is adopted to generate tree-level Feynman
diagrams and convert them to corresponding amplitudes. We present
the t-channel tree-level Feynman diagrams involving strong
interaction for the process \rrttbb in Fig.1. The u-channel
diagrams, which can be obtained by exchanging initial photons of
the corresponding t-channel ones, are not drawn there.
\begin{figure*}
\begin{center}
\includegraphics[width=10.5cm]{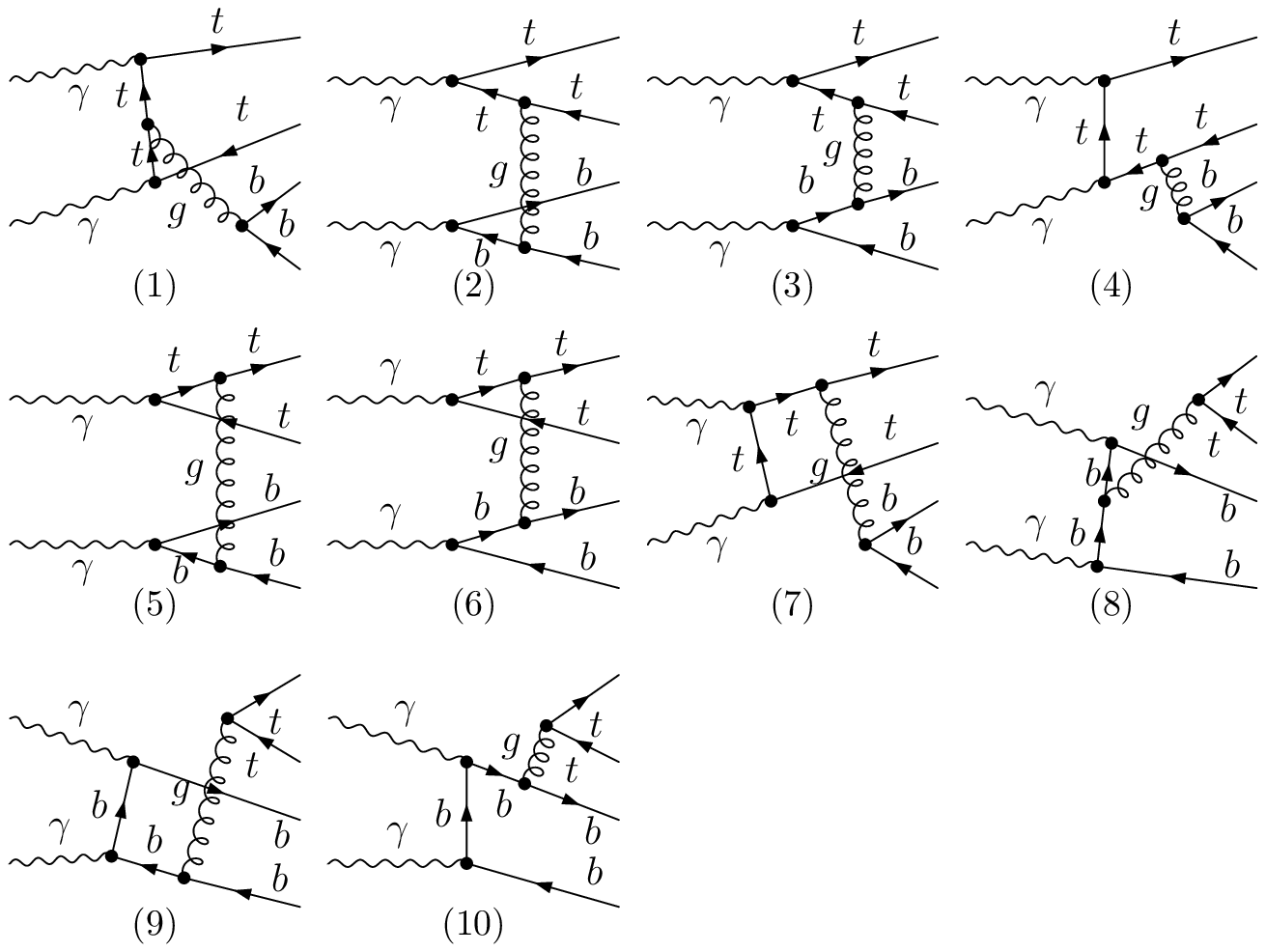}
\caption{\label{Fig1} The t-channel tree-level Feynman diagrams at
the ${\cal O}(\alpha_{ew}\alpha_s)$ order for \rrttbb. }
\end{center}
\end{figure*}

\par
The tree-level amplitude ${\cal M}$ of the process \rrttbb at the
${\cal O}(\alpha_{ew}\alpha_s)$ order, is then expressed as below.
\begin{equation} \label{Mat}
{\cal M}  = \sum_{i=1}^{10}\sum_{j=u}^{t}{\cal M}_j^{(i)}.
\end{equation}
where the amplitudes (${\cal M}_t^{(i)},~i=1,...,10$) correspond
to the diagrams in Fig.\ref{Fig1}(1-10) respectively. Then we get
the differential cross section for the process \rrttbb at the
tree-level as
\begin{equation}
d\sigma_{tree}= \overline{\sum}\left|{\cal M}\right|^2 {d\Phi_4},
\label{Sig}
\end{equation}
%--------------
where $d\Phi_4$ is the four-body phase space element. The
summation is taken over the spins and colors of final states, and
the bar over the summation in Eq.(\ref{Sig}) recalls averaging
over initial spin states. The calculation of the amplitudes of
tree-level diagrams for \rrttbb process is implemented by using
FormCalc4.1 package\cite{formloop}. The integration program of the
four-body phase space is based on the FormCalc4.1 package, and
created by using the factorization expression of four-body phase
space element\cite{monte},
\begin{equation}
d\Phi_4(p_1+p_2,p_3,p_4,p_5,p_6)=\frac{1}{2\pi}dQ^2d\Phi_3(p_1+p_2,p_3,p_4,Q)
d\Phi_2(Q,p_5,p_6).
\end{equation}
where $Q \equiv p_5+p_6$ and the actual integration of three-body
phase space element $d\Phi_3(p_1+p_2,p_3,p_4,Q)$ is parameterized
using 2to3.F program in FormCalc4.1 package\cite{formloop}, and
the complete integration over the four-body phase space is
performed using Monte Carlo integrator Vegas. The two-body phase
space $d\Phi_2(Q,p_5,p_6)$ is given as
\begin{equation}
d\Phi_2(Q,p_5,p_6)=\frac{1}{(2\pi)^2}\frac{\sqrt{\lambda(Q^2,m_5^2,m_6^2)}}
{8Q^2}d\varphi_5d(cos\theta_5),
\end{equation}
where the kinematical function $\lambda(x,y,z)$ is defined by
\begin{equation}\label{Lamb}
\lambda(x,y,z)=x^2+y^2+z^2-2xy-2yz-2zx.
\end{equation}

\par
In order to check our four-body phase space integration program,
we calculated the LO cross section of $e^+e^-\to u \bar d \mu^-
\bar \nu_{\mu}$ process by adopting fixed-width method and the
same input parameters as used in Ref.\cite{denner2}. In Table
\ref{tab1} we list the the LO cross sections obtained from
Ref.\cite{denner2}, by using GRACE2.2.0 system\cite{Grace}, and by
adopting our 4-body phase space integration program together with
FeynArts3.2 and FormCalc4.1, respectively. It demonstrates the
numerical integration results by using our created program are in
good agreement with others.
\begin{table}
\begin{center}
\begin{tabular}{|c|c|c|c|}
\hline $\sqrt{s}(GeV)$ & $\sigma_{LO}(fb)$(Ref.\cite{denner2})
& $\sigma_{LO}(fb)$(GRACE)  & $\sigma_{LO}(fb)$(ours)  \\
\hline 200 & 661.3(3) & 661.24(3) & 661.26(4) \\
500 & 260.9(1) & 260.85(2) & 260.87(2) \\
\hline
\end{tabular}
\end{center}
\begin{center}
\begin{minipage}{15cm}
\caption{\label{tab1} The comparison of the numerical results of
the LO cross section for $e^+e^-\to u \bar d \mu^- \bar \nu_{\mu}$
process by using GRACE2.2.0 system and our integration program
with those presented in Ref.\cite{denner2}. }
\end{minipage}
\end{center}
\end{table}

\vskip 10mm
\section{The NLO QCD corrections
to the process \rrttbb}
\par
\begin{figure}[htbp]
\begin{center}
\includegraphics[width=11.5cm]{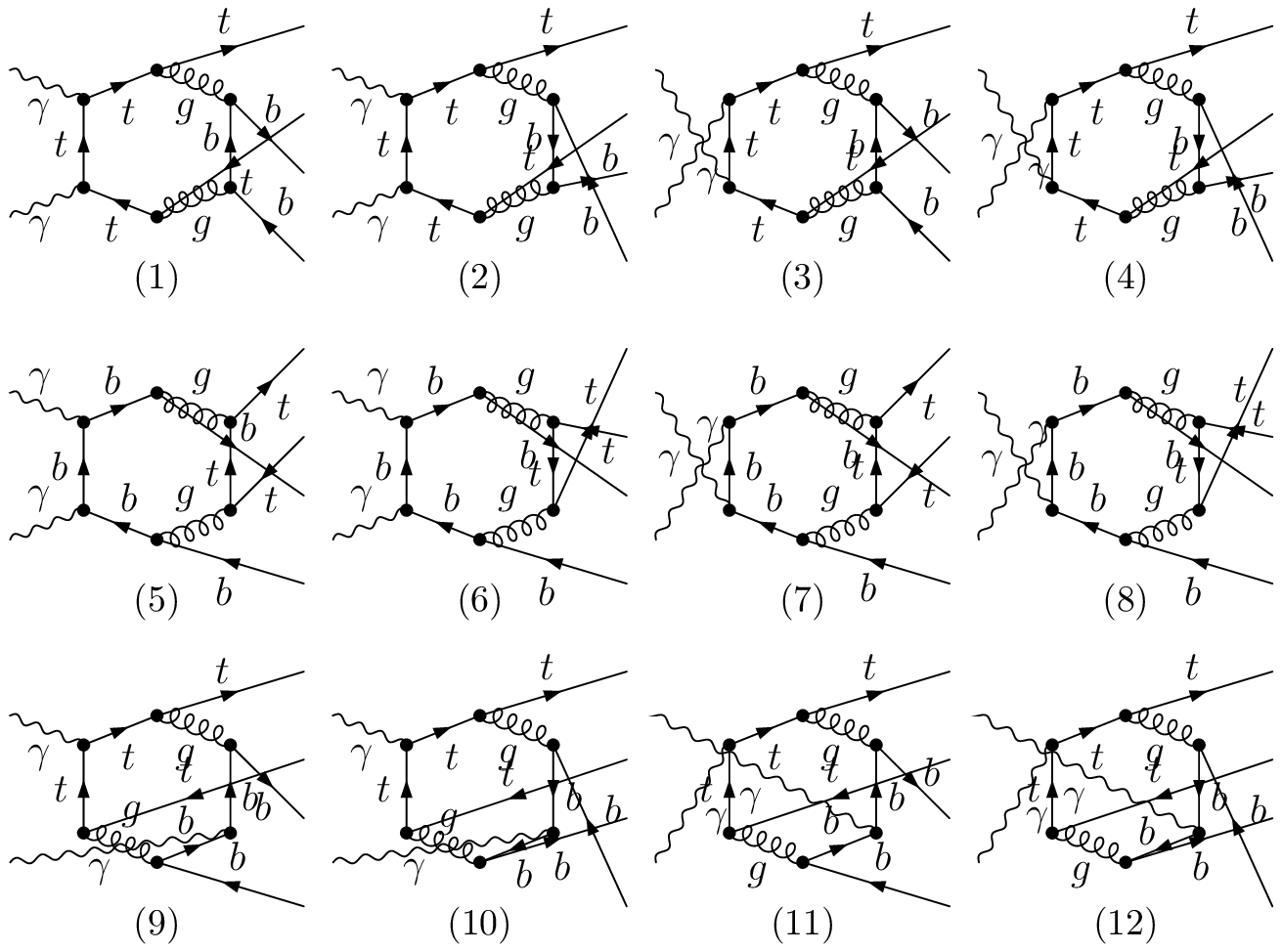}
\vspace*{-0.3cm} \centering \caption{\label{Fig2} The hexagon
Feynman diagrams for \rrttbb process. }
\end{center}
\end{figure}

The FeynArts3.2 package is used to generate QCD one-loop Feynman
diagrams of \rrttbb process at the order of ${\cal
O}(\alpha_{ew}\alpha_s^2)$, and then to convert them to
corresponding amplitudes. The QCD one-loop Feynman diagrams can be
classified into 140 self-energy diagrams, 172 triangle diagrams,
108 box diagrams, 48 pentagon diagrams and 12 hexagon diagrams. We
also use the FormCalc4.1 package\cite{formloop} to calculate the
amplitudes of one-loop Feynman diagrams. But the original
FormCalc4.1 package doesn't possess the function to calculate the
amplitudes including 6-point integrals, we have to create some
codes in FormCalc4.1 in order to handle amplitudes relevant to
hexagon diagrams. As a representative selection, we present the
hexagon Feynman diagrams of the \rrttbb process in Fig.\ref{Fig2}.
There exist both ultraviolet(UV) divergency and soft infrared(IR)
singularity in the contribution part of virtual gluon one-loop
diagrams for \rrttbb process, but no collinear IR singularity due
to the nonzero masses of top and bottom quark. Dimensional
regularization(DR) scheme in $D=4-2 \epsilon$ dimensions is
adopted to isolate both IR and UV singularities. By adopting
$\overline{MS}$-scheme to renormalize the strong coupling strength
and the $OS$-scheme to renormalize the masses and fields of top-
and bottom-quark, the UV singularities are vanished.
%%%%figure%%%%

\par
In the total cross section, the soft IR divergency contributed by
the virtual correction part should be cancelled by the
contribution from real gluon emission process \rrttbb$+g$ at
tree-level(${\cal O}(\alpha_{ew}^2\alpha_s^3)$), which presents
the same order contribution as the virtual correction does. The
real gluon emission process is denoted as
\begin{eqnarray}
\gamma(p_1)+\gamma(p_2)\to t(p_3)+\bar t(p_4)+b(p_5)+\bar
b(p_6)+g(p_7).
\end{eqnarray}

We adopt the two cutoff phase-space slicing(TCPSS)
method\cite{twocut} to calculate the real gluon emission process.
Since there is no collinear IR singularity, we introduce only an
arbitrary small soft cutoff $\delta_s$ to separate the \rrttbb$+g$
phase space into two regions, according to whether the energy of
the emitted gluon is soft($E_7 \leq \delta_s\sqrt{s}/2$) or
hard($E_7 > \delta_s\sqrt{s}/2$). After a lengthy calculation
similar with those shown in references
\cite{ppqcd2,Catani:1996jh,Catani:2002hc,ppqcd1}, we can get the
expression of $\sigma_{soft}$ for \rrttbb process as,
\begin{equation}
\sigma_{soft}=\frac{\alpha_s}{2\pi}\left[\frac{1}{3}\left(g_{35}+
g_{46}\right)+\frac{7}{6}\left(g_{36}+g_{45}\right)-
\frac{1}{6}\left(g_{34}+g_{56}\right)\right]\sigma_{tree},
\end{equation}
where $g_{ij}~(i,j=3,4,5,6)$ are defined as,
\begin{eqnarray}
g_{ij}(p_i,p_j) &=& \left(\frac{\pi\mu^2}{\Delta
E^2}\right)^\epsilon \Gamma(1+\epsilon)\,
\left[\frac{2(p_ip_j)}{\lambda^{1/2}(s_{ij},m_i^2,m_j^2)}\ln(\sigma_i\sigma_j)+2\right]
\frac{1}{\epsilon}-\frac{2(p_ip_j)}{\lambda^{1/2}(s_{ij},m_i^2,m_j^2)}
\nb \\ &&
 \times \left[\frac{1}{2}\ln^2(\sigma_i)+\frac{1}{2}\ln^2(\sigma_j)
+2Li_2\left(1-\sigma_i\right) +2Li_2\left(1-\sigma_j\right)
\right] \nb \\
&& -\frac{1}{\rho_i}\ln\sigma_i -\frac{1}{\rho_j}\ln\sigma_j+{\cal
O(\epsilon)}, ~~~~~~(i,j=3,4,5,6,~i<j).   \label{gij-2}
\end{eqnarray}
In above equation, $\lambda(s_{ij},m_i^2,m_j^2)$ is the
kinematical function defined in Eq.(\ref{Lamb}), $\Delta
E=\delta_s\sqrt{s}/2$, $s_{ij}=(p_i+p_j)^2$, $m_{3}=m_{4}=m_t$,
$m_{5}=m_{6}=m_b$ and
\begin{eqnarray}
\rho_i&=&\frac{\lambda^{1/2}(s_{ij},m_i^2,m_j^2)}{s_{ij}+m_i^2-m_j^2}, \nb \\
\sigma_i&=&\frac{1-\rho_i}{1+\rho_i}.
\end{eqnarray}

\par
Our calculation shows in the total cross section the soft IR
singularity induced by the one-loop virtual gluon correction is
exactly cancelled by the IR singularity part offered by the soft
gluon emission process \rrttbb$(g)$. The hard gluon emission cross
section $\sigma_{hard}$ for $E_7 > \delta_s\sqrt{s}/2$ is finite
and can be calculated numerically in four dimensions by using
Monte Carlo method. The integrations of 4-body phase space in
evaluations of $\sigma_{tree}$, $\sigma_{virtual}$ and
$\sigma_{soft}$ are carried out by using our created 4-body phase
space integration program as described in last section. But in
evaluation of $\sigma_{hard}$ of \rrttbb$+g$ process, we apply
CompHEP-4.4p3 program\cite{CompHEP} to calculate the tree-level
amplitude and the integration of 5-body phase space. Finally, the
NLO QCD corrected total cross sections for process \rrttbb can be
obtained as
\begin{eqnarray}
\sigma_{NLO}=\sigma_{tree}+\sigma_{virtual}+\sigma_{soft}+\sigma_{hard}.
\end{eqnarray}
The QCD NLO corrected cross section $\sigma_{NLO}$ is both UV- and
IR-finite. All the UV and IR divergences are cancelled
analytically.

\vskip 10mm
\section{Numerical Results and Discussions}
\par
In our numerical calculation we take following input
parameters\cite{hepdata,leger}:
\begin{equation}
\begin{array}{ll}
m_W=80.403~GeV,&~~~~m_Z=91.1876~GeV, \\
m_t=172.5~GeV,&~~~~m_b=4.7~GeV. \\
\alpha_{ew}(0)^{-1}=137.0359991,&~~~~\alpha_s(m_Z^2)=0.1176,
\end{array}
\end{equation}
The QCD renormalization scale $\mu$ is taken to be
$\mu=\mu_0(\equiv m_t+m_b)$ if there is no other statement, and
the running strong coupling $\alpha_s(\mu^2)$ is evaluated at the
three-loop level ($\overline{MS}$ scheme) with the five active
flavors\cite{hepdata}. For the numerical calculations of one-loop
integrals, we use LoopTools2.1\cite{formloop} package to deal with
2-, 3- and 4-point integrals. The implementations of the scalar
and tensor 5-point integrals are done exactly by using our Fortran
programs as used in previous works\cite{youyu,ZhangRY} with the
approach presented in Ref.\cite{pentagon}. And the 6-point scalar
and tensor integrals are evaluated by using our created programs
with the expressions given in Refs.\cite{denner1} and
\cite{denner3}. In Ref.\cite{Binoth1}, T. Binoth, {et al.,}
derived an analytic expression for the scalar hexagon function,
which is convenient for the subsequent numerical integration. We
checked also the numerical results of 6-point scalar integrals by
using two methods presented in Refs.\cite{denner1,denner3} and
\cite{Binoth1}, and confirmed the correctness of our 6-point
scalar integral program. For example, with the Set(I) input
parameters in Ref.\cite{Binoth1} we get exactly the same numerical
results as $(1.3526\times 10^{-2}+4.0608\times 10^{-15}~i)$ by
using both two methods.

\par
\begin{figure}[htp]
\vspace*{-0.3cm} \centering
\includegraphics[scale=0.36]{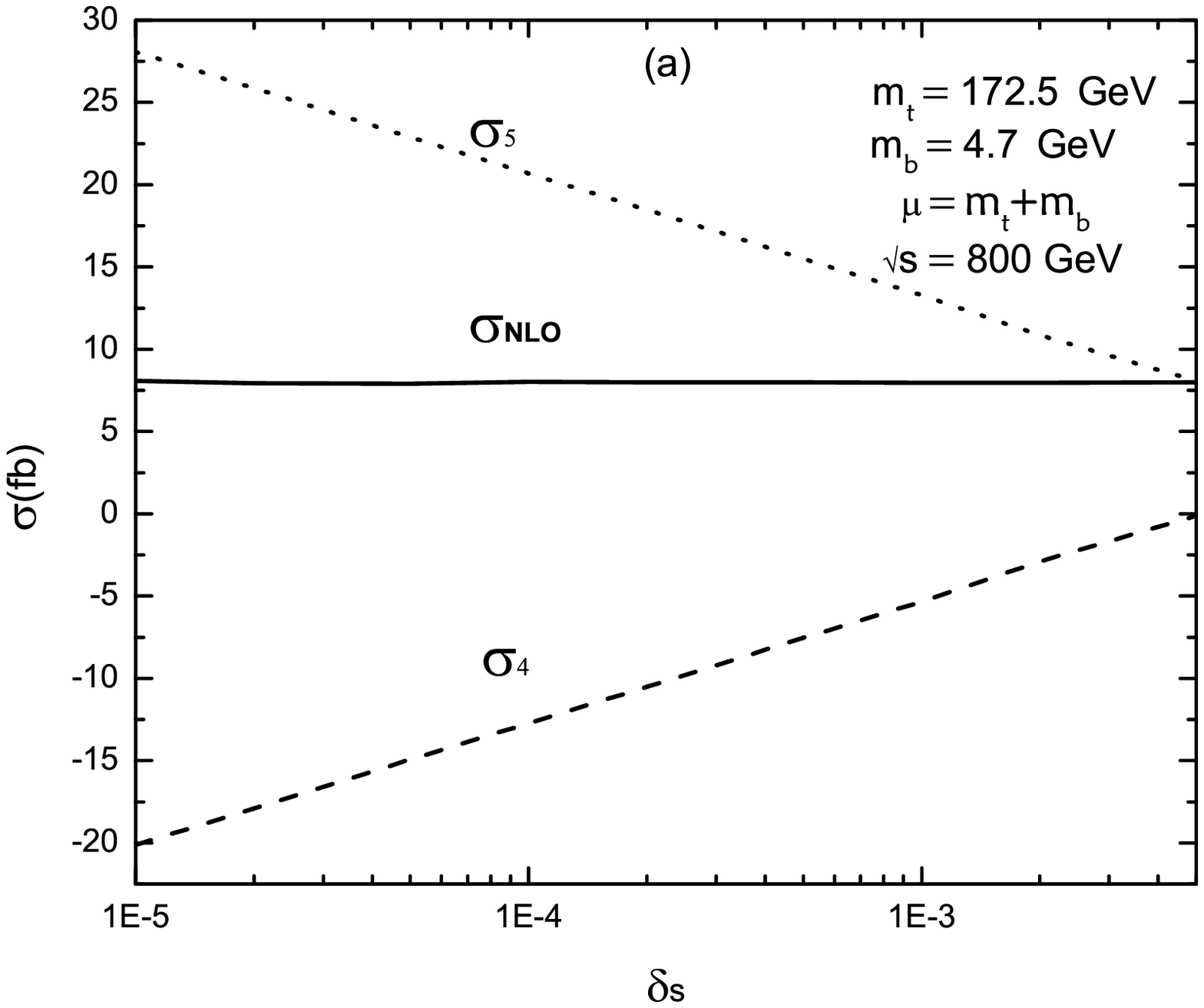}
\includegraphics[scale=0.36]{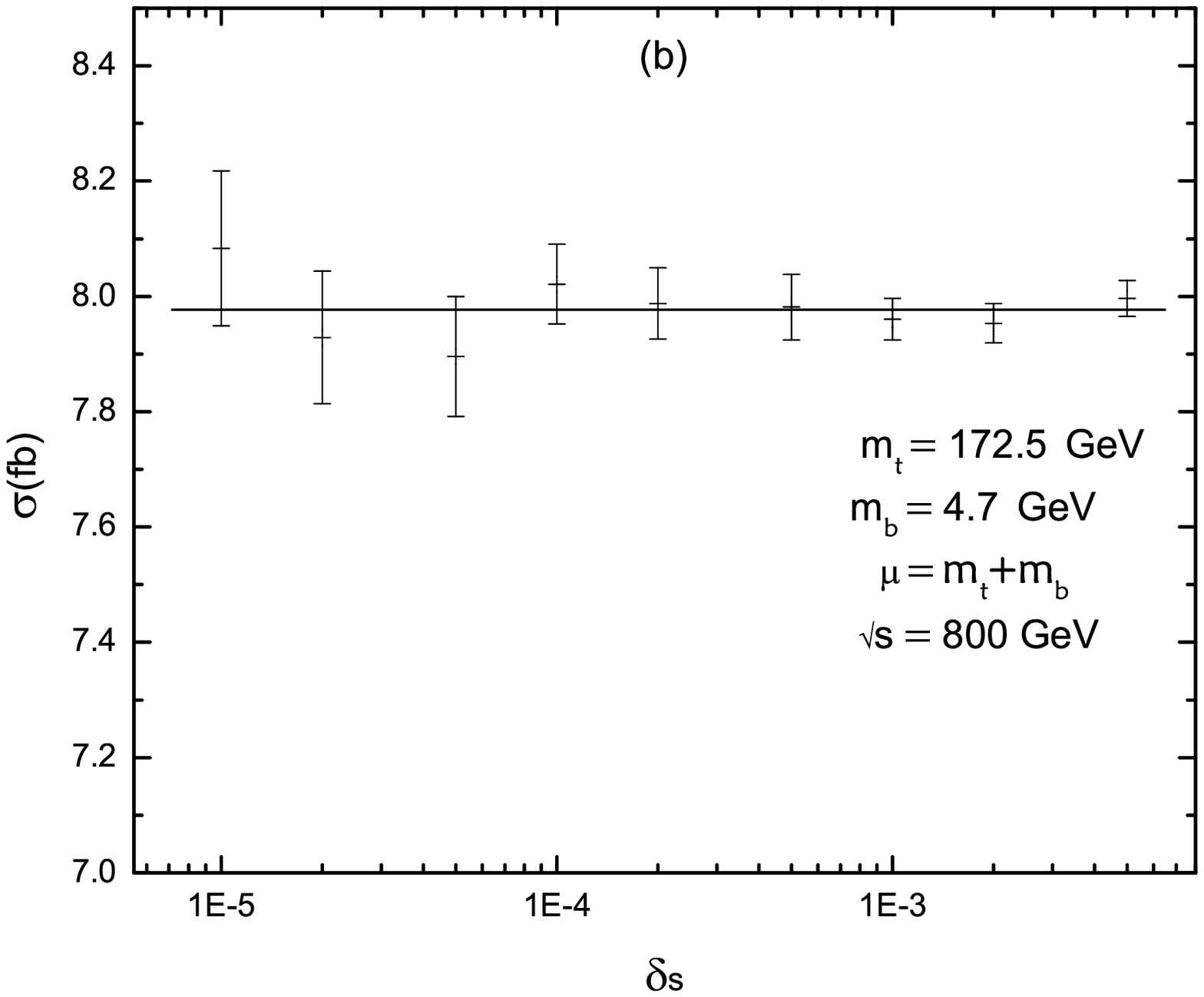}
\caption{\label{fig3ab} (a) The cross section parts
$\sigma_4(=\sigma_{tree}+\sigma_{soft}+\sigma_{virtual})$,
$\sigma_5(=\sigma_{hard})$ and the NLO QCD corrected cross
section($\sigma_{NLO}$) of the \rrttbb process as the functions of
the soft cutoff $\delta_s(=2~\Delta E_7/\sqrt{s})$ with
$\mu=\mu_0=m_t+m_b$ and $\sqrt{s}=800~{\rm GeV}$. (b) The enlarged
plot of Fig.\ref{fig3ab}(a) for the NLO QCD corrected cross
section($\sigma_{NLO}$) with integration error versus $\delta_s$.
}
\end{figure}

During our numerical calculation, we have studied the independence
of the total cross section involving the NLO QCD corrections of
process \rrttbb on the soft cutoff $\delta_s(=2~\Delta
E_7/\sqrt{s})$. To show that independence, we depict the cross
section parts
$\sigma_4(=\sigma_{tree}+\sigma_{soft}+\sigma_{virtual})$,
$\sigma_5(=\sigma_{hard})$ and NLO QCD corrected cross section
$\sigma_{NLO}$ versus $\delta_s$ in Figs.\ref{fig3ab}(a-b) with
$\sqrt{s}=800~{\rm GeV}$. As shown in these two figures, both
$\sigma_4$ and $\sigma_5$ obviously depend on the soft cutoff
$\delta_s$, but $\sigma_{NLO}$ is independent of the soft cutoff
value with the best fit average value being $7.978~fb$ and the
errors being less than $1.3\%$ and $0.5\%$ in the $\delta_s$
regions of $[10^{-5},~10^{-4}]$ and $[10^{-4},~5\times 10^{-2}]$,
respectively. In further numerical calculation we fix
$\delta_s=10^{-3}$.
%%figure%%
\vskip 10mm

\par
\begin{figure}
\includegraphics[scale=0.36]{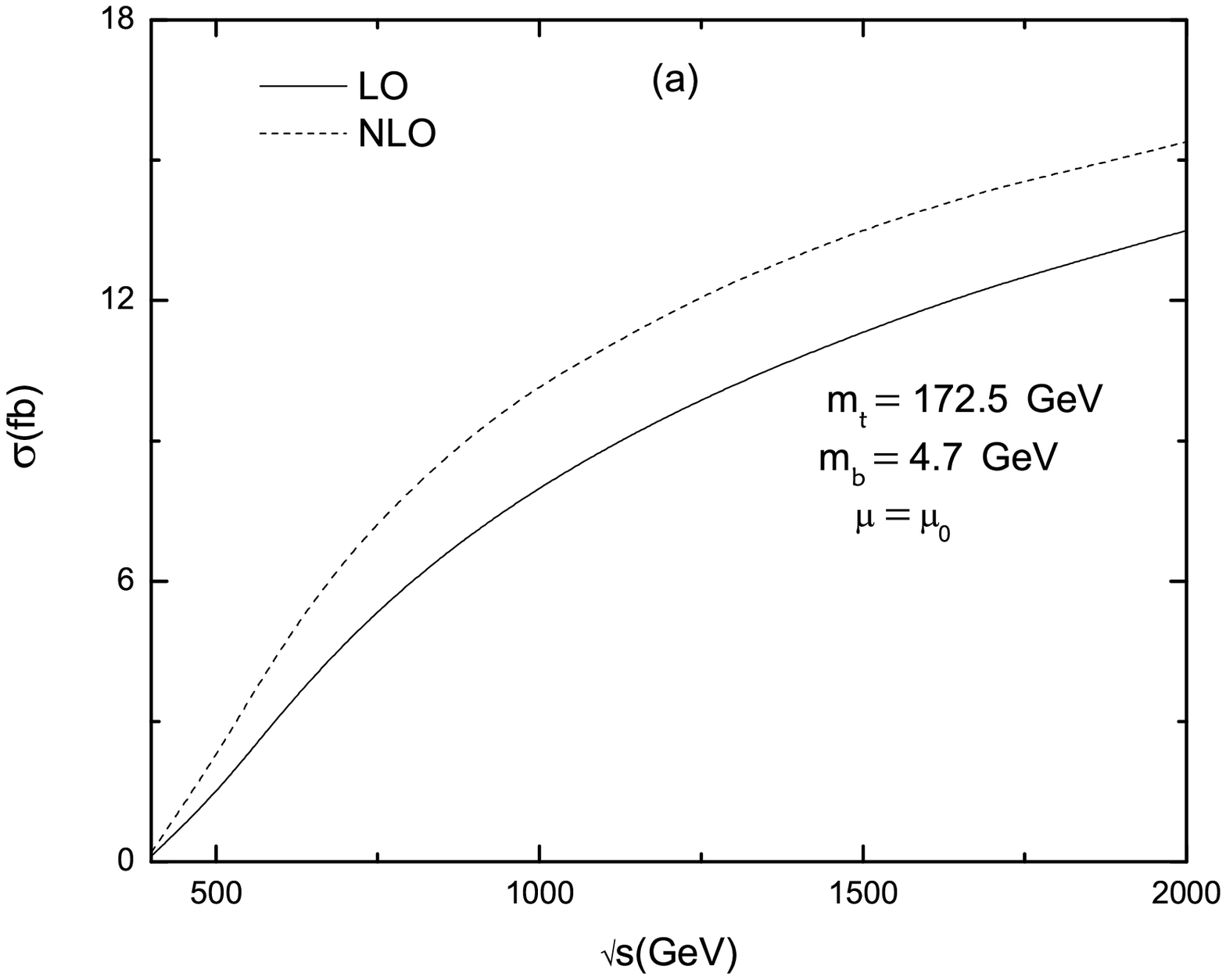}
\includegraphics[scale=0.36]{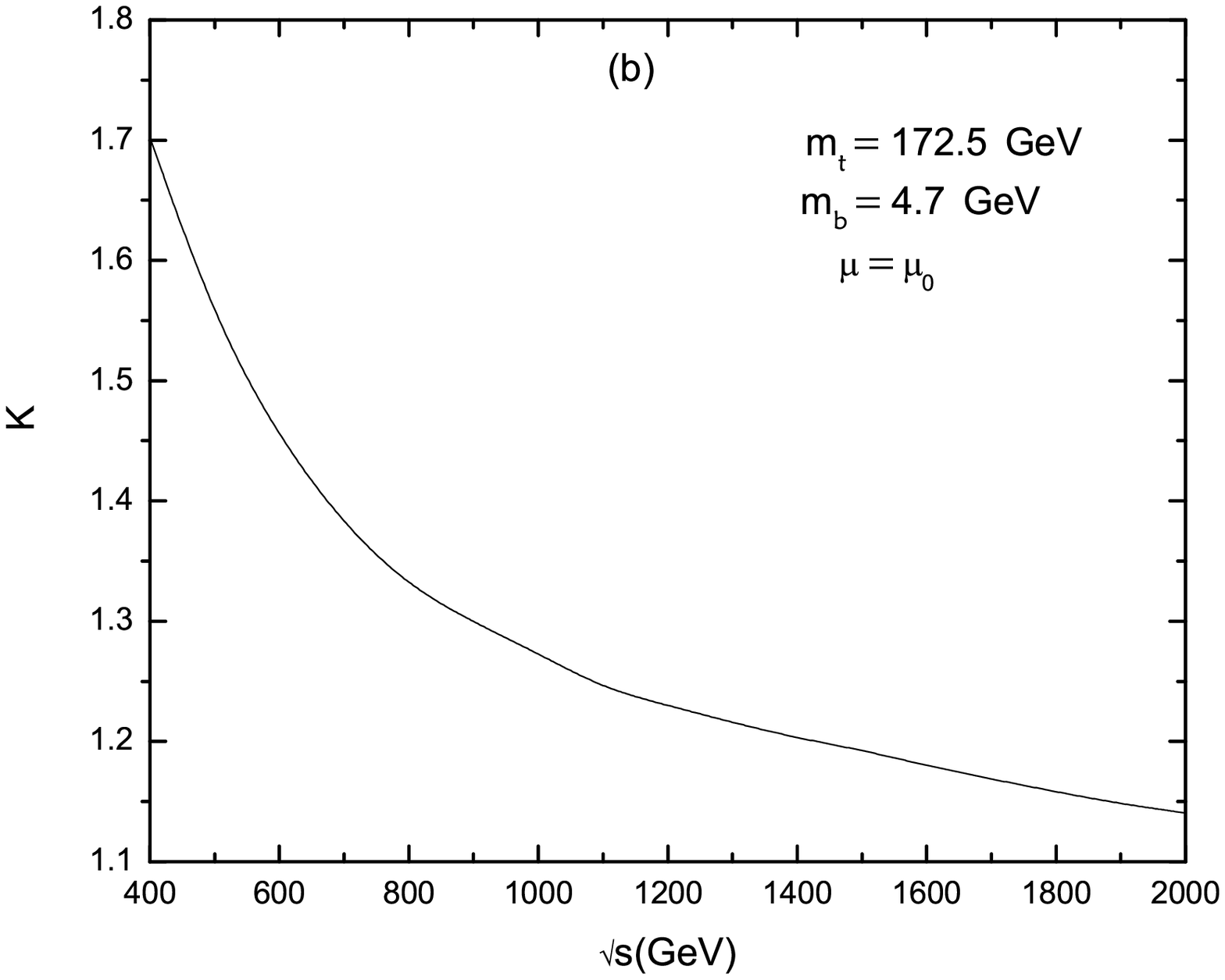}
\caption{\label{fig3} (a) The LO and NLO QCD corrected cross
sections for the process \rrttbb as the functions of c.m.s.
colliding energy($\sqrt{s}$), (b) the corresponding K-factor
versus $\sqrt{s}$.}
\end{figure}

We define the K-factor as the ratio of the NLO QCD corrected cross
section and the LO cross section($K \equiv\frac{\sigma_{NLO}}
{\sigma_{LO}}$). The numerical results of the cross section and
the K-factor for the process \rrttbb are plotted in
Fig.\ref{fig3}(a) and Fig.\ref{fig3}(b) respectively, with the
$\gamma\gamma$ colliding energy $\sqrt{s}$ running from $400~GeV$
to $2~TeV$. In Fig.\ref{fig3}(a), the full and the dashed curves
correspond to the LO and NLO QCD corrected cross sections
separately. As indicated in Fig.\ref{fig3}(a), the cross section
increases quickly as the $\gamma\gamma$ colliding energy running
from $400~GeV$ to $2~TeV$, and the NLO QCD correction obviously
enhances the tree-level cross section in the plotted range of
$\sqrt{s}$. The NLO QCD corrected cross section of process \rrttbb
with $\sqrt{s}=2~TeV$ can reach the value of $15.39~fb$.
Fig.\ref{fig3}(b) shows the corresponding K-factor varies from
$1.70$ to $1.14$ as the c.m.s energy $\sqrt{s}$ running from
$400~GeV$ to $2~TeV$. The analysis of the contribution parts of
NLO QCD correction shows that in the small energy region the
K-factor is enhanced due to a Coulomb singularity effect on the
contribution from the diagrams with virtual gluon exchange between
heavy quarks. The cross sections, $\sigma_{tree}$ and
$\sigma_{NLO}$, and K-factors at some typical $\sqrt{s}$ points
can be read out from Figs.\ref{fig3}(a-b) and are listed in Table
\ref{tab2}. For the correctness check of the calculation of the LO
cross section for \rrttbb process, the results obtained by using
GRACE2.2.0 system are presented there too.

\begin{table}
\begin{center}
\begin{tabular}{|c|c|c|c|c|}
\hline
$\sqrt{s}(GeV)$ & $\sigma_{tree}(fb)$(GRACE) & $\sigma_{tree}(fb)$
& $\sigma_{NLO}(fb)$ & K-factor \\
\hline
500 & 1.4453(4) & 1.4458(5)  & 2.24(1)  & 1.55  \\
800 & 5.991(3)  & 5.990(4)   & 7.96(4)  & 1.33  \\
1000 & 8.001(6) & 8.000(7)   & 10.18(6) & 1.27  \\
2000 & 13.51(2) & 13.50(2)   & 15.39(9) & 1.14  \\
\hline
\end{tabular}
\end{center}
\begin{center}
\begin{minipage}{15cm}
\caption{\label{tab2} The LO and NLO QCD corrected cross sections,
K-factors for \rrttbb process with $\sqrt{s}=500~GeV$, $800~GeV$,
$1000~GeV$, $2000~GeV$, respectively. The LO cross section values
obtained by using GRACE2.2.0 system are listed there for
comparison. The number of the Monte Carlo events is $5\times
10^6$.}
\end{minipage}
\end{center}
\end{table}

\par
\begin{figure}
\begin{center}
\includegraphics[scale=0.36]{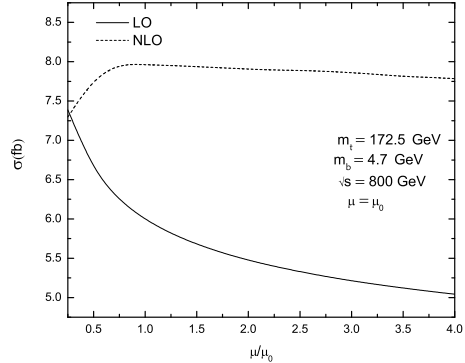}
\caption{\label{fig4} The LO and NLO QCD corrected cross sections
for the \rrttbb process as the functions of renormalization scale
$\mu/\mu_0(\mu_0 \equiv m_t+m_b=147.2~GeV)$.}
\end{center}
\end{figure}

The renormalization scale dependence of both the LO and NLO QCD
corrected total cross sections for the process \rrttbb with
$\sqrt{s}=800~GeV$, is plotted in Fig.\ref{fig4}. In this figure
the scale is parameterized as $\mu/\mu_0~(\mu_0 \equiv m_t+m_b)$
and the full-line and dashed-line correspond to the LO and the NLO
QCD corrected cross sections, respectively. We can see from the
figure that in the region $0.75 < \mu/\mu_0 < 4$ the NLO QCD
correction obviously improves the independence of the
renormalization scale $\mu$. Therefore, we can conclude that the
uncertainty of the cross section for process \rrttbb due to the
variation of renormalization scale $\mu$, can be reduced by
considering the NLO QCD corrections.

\par
\begin{figure}
\includegraphics[scale=0.36]{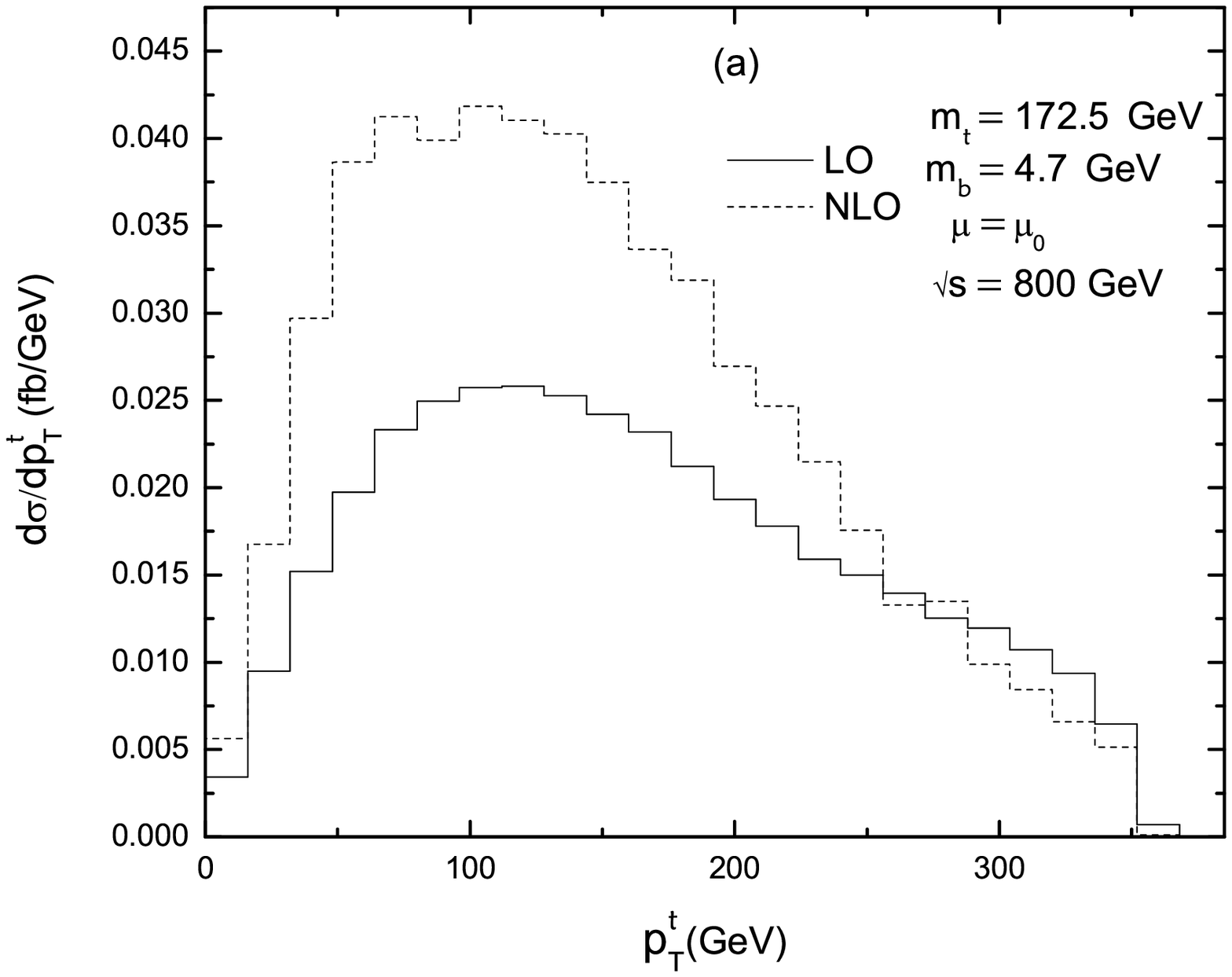}
\includegraphics[scale=0.36]{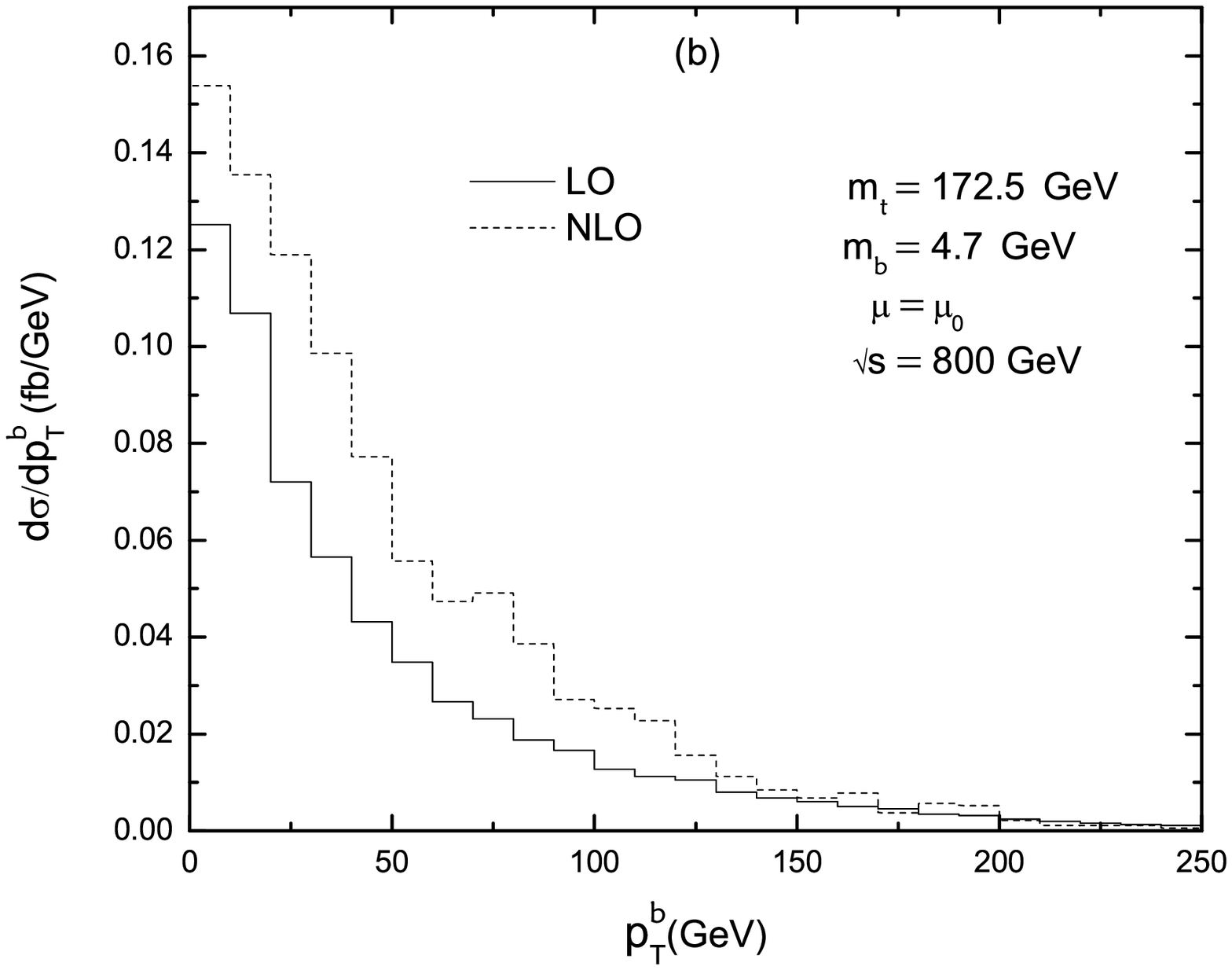}
\caption{\label{fig5} The distributions of the transverse momenta
of top- and bottom-quark for the \rrttbb process with
$\sqrt{s}=800~GeV$. (a) for top-quark, (b) for bottom-quark.}
\end{figure}

The distributions of the transverse momenta of top- and
bottom-quark($p_T^t$ and $p_T^b$) with the colliding energy
$\sqrt{s}=800~GeV$, are depicted in Fig.\ref{fig5}(a) and
Fig.\ref{fig5}(b) separately. In Fig.\ref{fig5}(a), we can see
that the NLO QCD correction obviously enhances the differential
cross section predicted in the $SM$ at tree-level when the $p_T^t$
value is less than $250~GeV$. Our analysis shows that in the
region $20~GeV<p_T^t<250~GeV$, the large correction to the
distribution of $p_T^t$ comes mainly from the contribution of the
hard gluon emission process. But when $p_T^t>250~GeV$, the
correction to the distribution of $p_T^t$ from the hard gluon
emission process, is largely cancelled by the negative correction
from virtual gluon and soft gluon emission contributions, then the
$p_T^t$ distribution corrections become to be much smaller. That
means the absolute correction
($|\frac{d\sigma_{NLO}}{dp_T^t}-\frac{d\sigma_{tree}}{dp_T^t}|$)
is enhanced at low $p_T^t$ and reduced at large $p_T^t$ due to the
momentum balance between top particles and gluons radiated from
top quarks at the NLO, which reduces the momenta of the top
quarks. For b-quark, we can see from Fig.\ref{fig5}(b) that the
large enhancement of the QCD corrected differential cross
section($\frac{d\sigma_{NLO}}{dp_T^b}$), which can nearly double
the LO differential cross section somewhere, can be appeared in
the $p_T^b$ value range between $20~GeV$ and $120~GeV$, while in
the range of $p_T^b>150~GeV$ the NLO QCD correction becomes to be
very small. Similar with the discussion for the distribution of
$p_T^t$, that is also the consequence of the momentum balance
between bottom particles and gluons radiated from bottom quarks at
the NLO.

\vskip 10mm
\section{Summary}
\par
In this paper we calculate the complete one-loop QCD corrections
to the process \rrttbb at a photon-photon collider. We present the
dependence of the NLO QCD correction of process \rrttbb on
colliding energy $\sqrt{s}$ in the $SM$, and find that NLO QCD
correction can generally increase the LO cross section. It shows
that in the $\gamma\gamma$ colliding energy range of
$400~GeV<\sqrt{s}<2~TeV$, the corresponding K-factor goes down
from $1.70$ to $1.14$. We find that the NLO QCD correction can
obviously improve the independence of the cross section for
process \rrttbb on the renormalization scale $\mu$, and the NLO
QCD correction also changes obviously the distributions of
transverse momenta of the final top- and bottom-quark states.

\vskip 3mm
\par
\noindent{\large\bf Acknowledgments:} We would like to acknowledge
professor C.-S. Li for bringing our attention to this issue. This
work was supported in part by the National Natural Science
Foundation of China, the Education Ministry of China and a special
fund sponsored by Chinese Academy of Sciences.

\end{document}